\documentclass{PoS}
\usepackage{subfigure}

\title{Chemical abundances in metal-poor giants: limitations imposed by the use of classical 1D stellar atmosphere models}

\ShortTitle{Chemical abundances in metal-poor giants}

\author{\speaker{V. Dobrovolskas}\\
        Vilnius University Astronomical Observatory, Vilnius, Lithuania\\
        E-mail: \email{vidas.dobrovolskas@ff.vu.lt}}

\author{A. Ku\v{c}inskas\(^{ab}\)\\
        \llap{\(^a\)} Institute of Theoretical Physics and Astronomy of Vilnius University, Vilnius, Lithuania\\
        \llap{\(^b\)} Vilnius University Astronomical Observatory, Vilnius, Lithuania\\
        E-mail: \email{ak@itpa.lt}}

\author{H.-G. Ludwig\(^{ab}\)\\
       \llap{\(^a\)}Landessternwarte Königstuhl, Heidelberg, Germany\\
       \llap{\(^b\)} GEPI, Observatoire de Paris, CNRS, Université Paris Diderot, France\\
       E-mail: \email{hludwig@lsw.uni-heidelberg.de}}

\author{E. Caffau\(^{ab}\)\\
       \llap{\(^a\)} Landessternwarte Königstuhl, Heidelberg, Germany\\
       \llap{\(^b\)} GEPI, Observatoire de Paris, CNRS, Université Paris Diderot, France\\
       E-mail: \email{Elisabetta.Caffau@obspm.fr}}

\author{J. Klevas, D. Prakapavi\v{c}ius\\
       Institute of Theoretical Physics and Astronomy of Vilnius University, Vilnius, Lithuania\\
       E-mail: \email{jonas.klevas@gmail.com, Dainius.Prakapavicius@tfai.vu.lt}}

\abstract{In this work we have used 3D hydrodynamical (CO\textsuperscript{5}BOLD) and 1D hydrostatic (LHD) stellar atmosphere models to study the importance of convection and horizontal temperature inhomogeneities in stellar abundance work related to late-type giants. We have found that for a number of key elements, such as Na, Mg, Si, Ca, Ti, Fe, Ni, Zn, Ba, Eu, differences in abundances predicted by 3D and 1D models are typically minor (< 0.1 dex) at solar metallicity. However, at [M/H]~=~--3 they become larger and reach to --0.5\,\dots--0.8 dex. In case of neutral atoms and fixed metallicity, the largest abundance differences were obtained for the spectral lines with lowest excitation potential, while for ionized species  the largest 3D--1D abundance differences were found for lines of highest excitation potential. The large abundance differences at low metallicity are caused by large horizontal temperature fluctuations and lower mean temperature in the outer layers of the 3D hydrodynamical model compared with its 1D counterpart.
}

\FullConference{11th Symposium on Nuclei in the Cosmos, NIC XI\\
		July 19-23, 2010\\
		Heidelberg, Germany}

\begin{document}

\section{Introduction}

Red giants are important tracers of intermediate age and old stellar populations and due to their high luminosity they are available for spectroscopic study beyond the Milky Way. Their abundances are generally determined using stationary 1D stellar atmosphere models which treat convection in a parametric way (e.g. microturbulence velocity, mixing length). These shortcomings of the classical 1D models are overcome in 3D hydrodynamic stellar atmosphere models which treat convection by solving time dependent hydrodynamic equations. Due to different approaches in the  treatment of convection one may expect differences in the predicted strengths of spectral lines, and thus -- differences in elemental abundances derived with the two types of models.

Comparison of abundance differences arising between the predictions of 3D and 1D stellar atmosphere models for red giants was made by [4]. Here we extended their work and used the 3D hydrodynamic stellar atmosphere models to investigate the influence of convection and horizontal temperature inhomogeneities on the predicted spectral line strength and thus -- on the derived elemental abundances in red giant stars. For this purpose we utilized a broader list of chemical elements and spectral line parameters, such as wavelength and excitation potential of the lower level, \(\chi\). This analysis was performed by means of the 3D--1D abundance corrections defined as an abundance difference between the 3D and 1D curves of growth obtained at a given equivalent width. The concept of the 3D--1D abundance correction was described in more detail in [3].

\section{Models}

In this work we used 3D hydrodynamic and 1D hydrostatic models calculated using the CO\textsuperscript{5}BOLD\footnote{http://www.astro.uu.se/\textasciitilde bf/co5bold\_main.html} [5,6] and LHD [2] codes, respectively. Both 3D and 1D models shared identical atmospheric parameters (Table 1), chemical composition, equation of state and opacities. Solar scaled chemical composition with alpha-element enhancement of +0.4 dex for metallicities [M/H]~\footnote{[M/H] = log[N(M)/N(H)]\(_\star\) - log[N(M)/N(H)]\(_\odot\).}~\( \leq \)~--1 is assumed in both 3D and 1D models [7]. Temperature stratifications of 3D and 1D models for [M/H] = 0 and --3 used in the current study are shown in Fig. 1.

\begin{table}[h]
\centering
\begin{tabular}{c c c c c}
\hline
\textit{T}\(_\mathrm{eff}\), K & log \textit{g} [cgs] & [M/H] & Grid dimension, Mm & Resolution\\
               &             &       & x\(\times\)y\(\times\)z & x\(\times\)y\(\times\)z\\
\hline
 4970 & 2.5 &   0    & 573\(\times\)573\(\times\)243 & 160\(\times\)160\(\times\)200\\
 4990 & 2.5 &  --1   & 573\(\times\)573\(\times\)245 & 160\(\times\)160\(\times\)200\\
 5020 & 2.5 &  --2   & 584\(\times\)584\(\times\)245 & 160\(\times\)160\(\times\)200\\
 5020 & 2.5 &  --3   & 573\(\times\)573\(\times\)245 & 160\(\times\)160\(\times\)200\\
\hline
\end{tabular}
\caption{Atmospheric parameters of the CO\textsuperscript{5}BOLD models used in this work.}
\label{tab1}
\end{table}

\begin{figure}[h]
\centering
\mbox{
      \subfigure{\includegraphics[width=7.2cm]{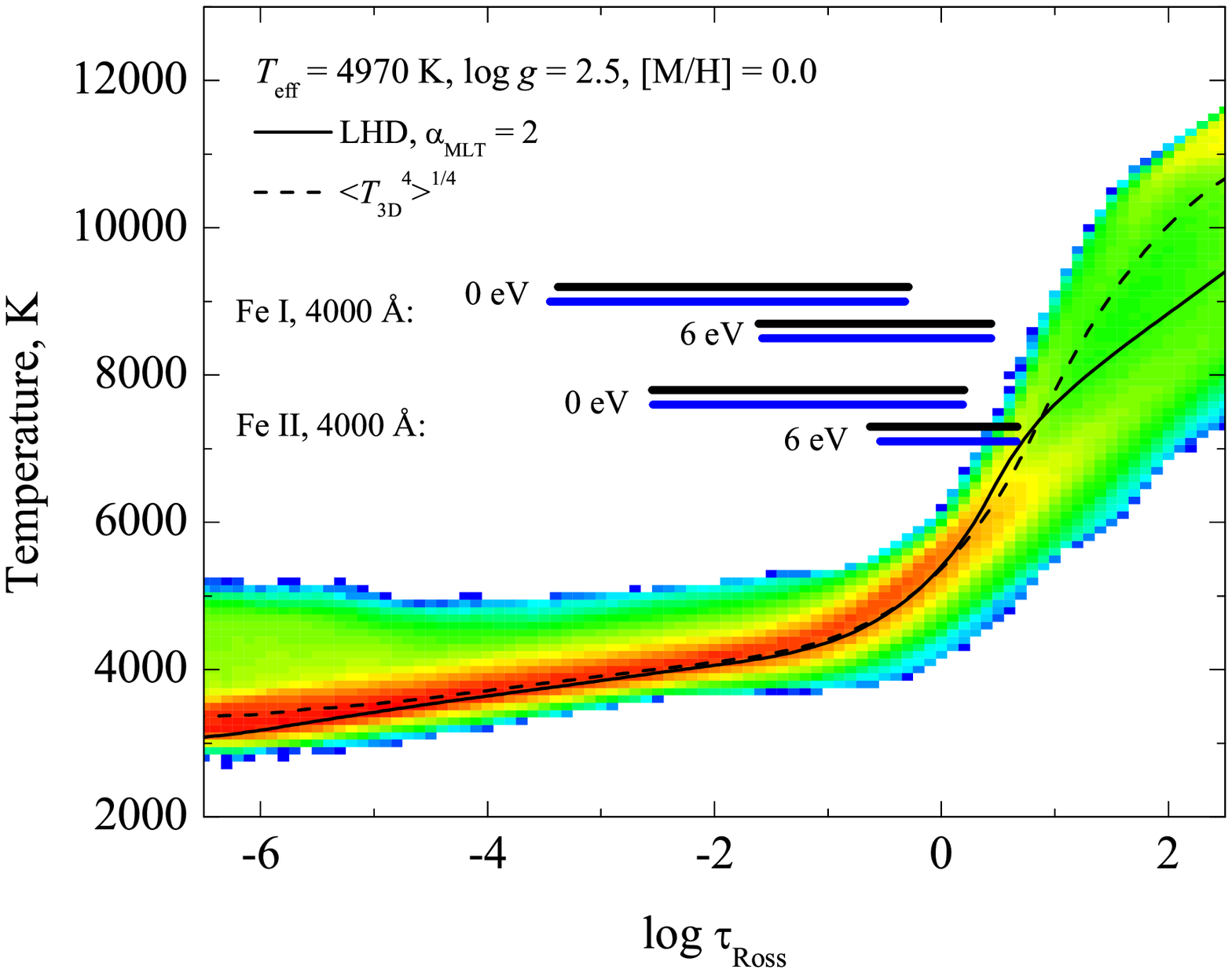}}
      \subfigure{\includegraphics[width=7.2cm]{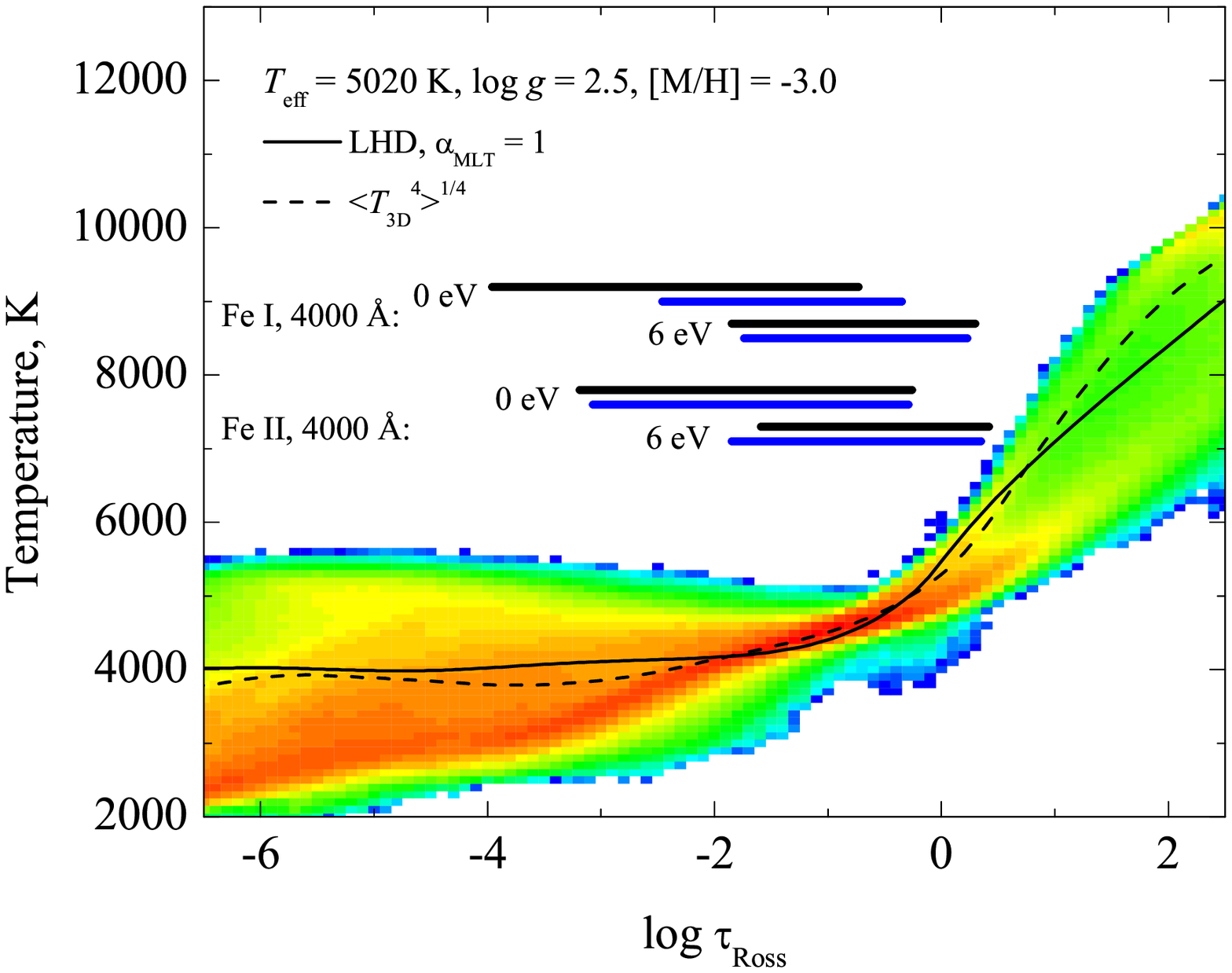}}
     }
\caption{Temperature profiles in two late-type giant models plotted versus Rosseland optical depth (left: \textit{T}\(_\mathrm{eff}\)/log \textit{g}/[M/H] = 4970/2.5/0.0; right: 5020/2.5/--3.0). The temperature distribution in the 3D model is shown with color-coded temperature increasing from blue to red. The dashed line shows the mean temperature of the 3D model averaged on surfaces of equal optical depth. The solid line is the temperature profile of the 1D model. Horizontal bars mark the regions where 90\% of the equivalent width of the Fe I and Fe II spectral line at \( \lambda \) = 4000~\AA{} and \(\chi\) = 0 eV and 6 eV is forming in the 3D (black) and 1D (blue) atmosphere models.}
\label{atm_mm00}
\end{figure}

\section{Spectral line synthesis}

Spectral line synthesis was done using the LINFOR3D\footnote{http://www.aip.de/\textasciitilde mst/linfor3D\_main.html} code which solves the 3D radiative transfer problem under the assumption of LTE.

The 3D--1D abundance corrections were calculated for a number of neutral and singly ionized elements: Na I, Mg I, Mg II, Si I, Si II, Ca I, Ca II, Ti I, Ti II, Fe I, Fe II, Ni I, Ni II, Zn~I, Zn II, Ba~II and Eu II.

Only weak lines (W < 5 m\AA) were used in calculations in order to avoid the influence of microturbulence parameter on the strength of 1D model lines. To investigate the dependence of 3D--1D abundance correction on wavelength and excitation potential we have synthesized fictitious lines at \( \lambda \) = 4000~\AA{} and 8500~\AA{} and excitation potentials in steps of 2 eV, from 0 to 6 eV.

\section{Results}

\subsection{Abundance corrections for neutral atoms}

The dependence of the 3D--1D abundance corrections on metallicity for neutral atoms is shown in Fig. 2 and 3, at \( \lambda \) = 4000 \AA{} and 8500 \AA, respectively. At solar metallicity the abundance corrections are small for all investigated chemical elements, never exceeding \(\pm\)0.05 dex at both wavelengths. All elements show increasingly more negative 3D--1D corrections with decreasing metallicity, larger in magnitude at \( \lambda \) = 8500~\AA{}. They are the largest in the case of magnesium and iron, and are approximately equal to --0.80 dex at [M/H]~= --3.0 for \( \lambda \) = 8500~\AA{} and the excitation potential of 0 eV (Fig. 2).

\begin{figure}[ht]
 \begin{minipage}[t]{0.5\linewidth}
 \centering
 \includegraphics[width=6.9cm]{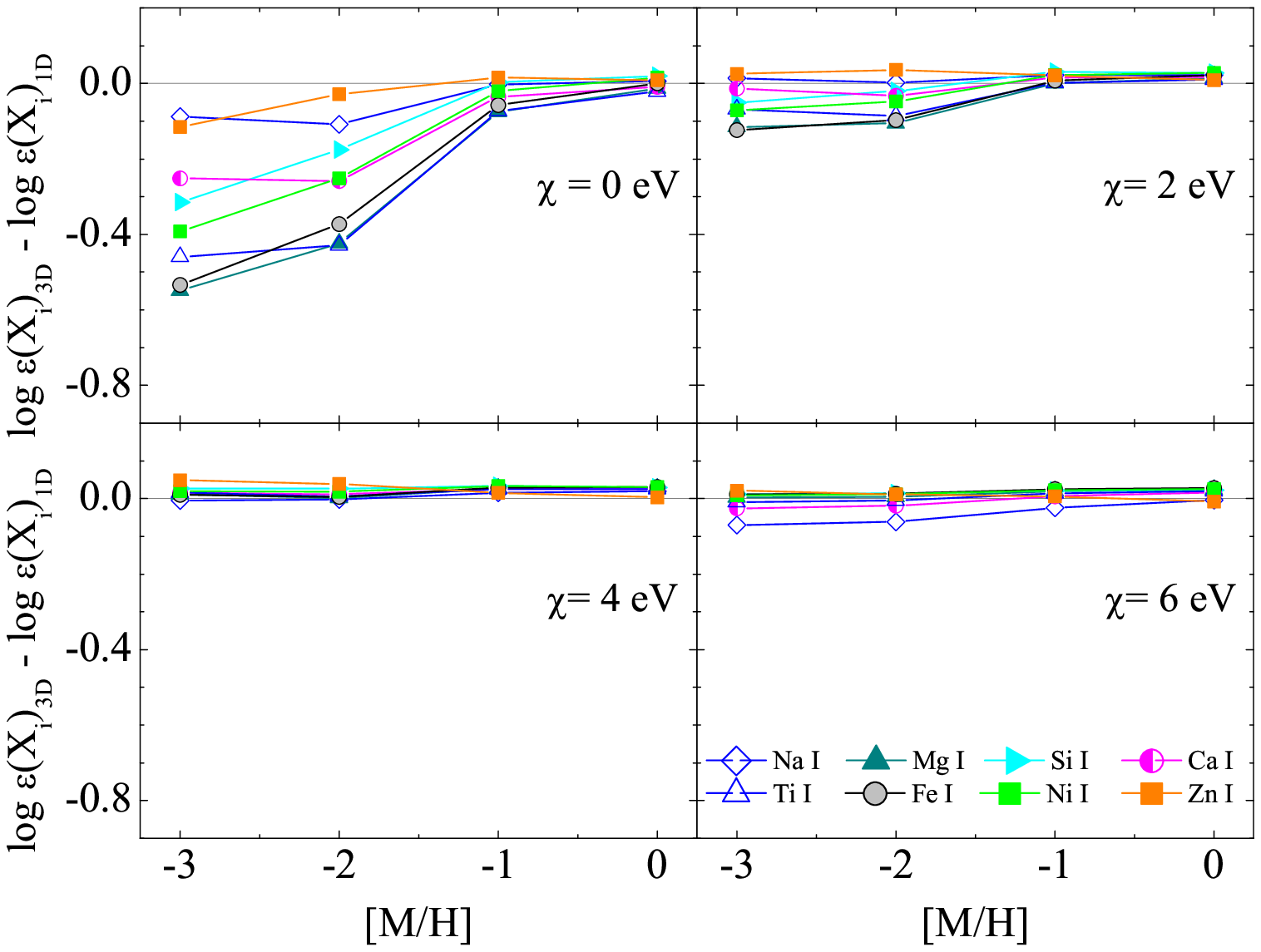}
 \caption{3D--1D abundance corrections for neutral atoms plotted as function of metallicity and excitation potential at \(\lambda\) = 4000 \AA.}
 \label{Fig2a}
 \end{minipage}
\hspace{0.0cm}
 \begin{minipage}[t]{0.5\linewidth}
 \centering
 \includegraphics[width=6.9cm]{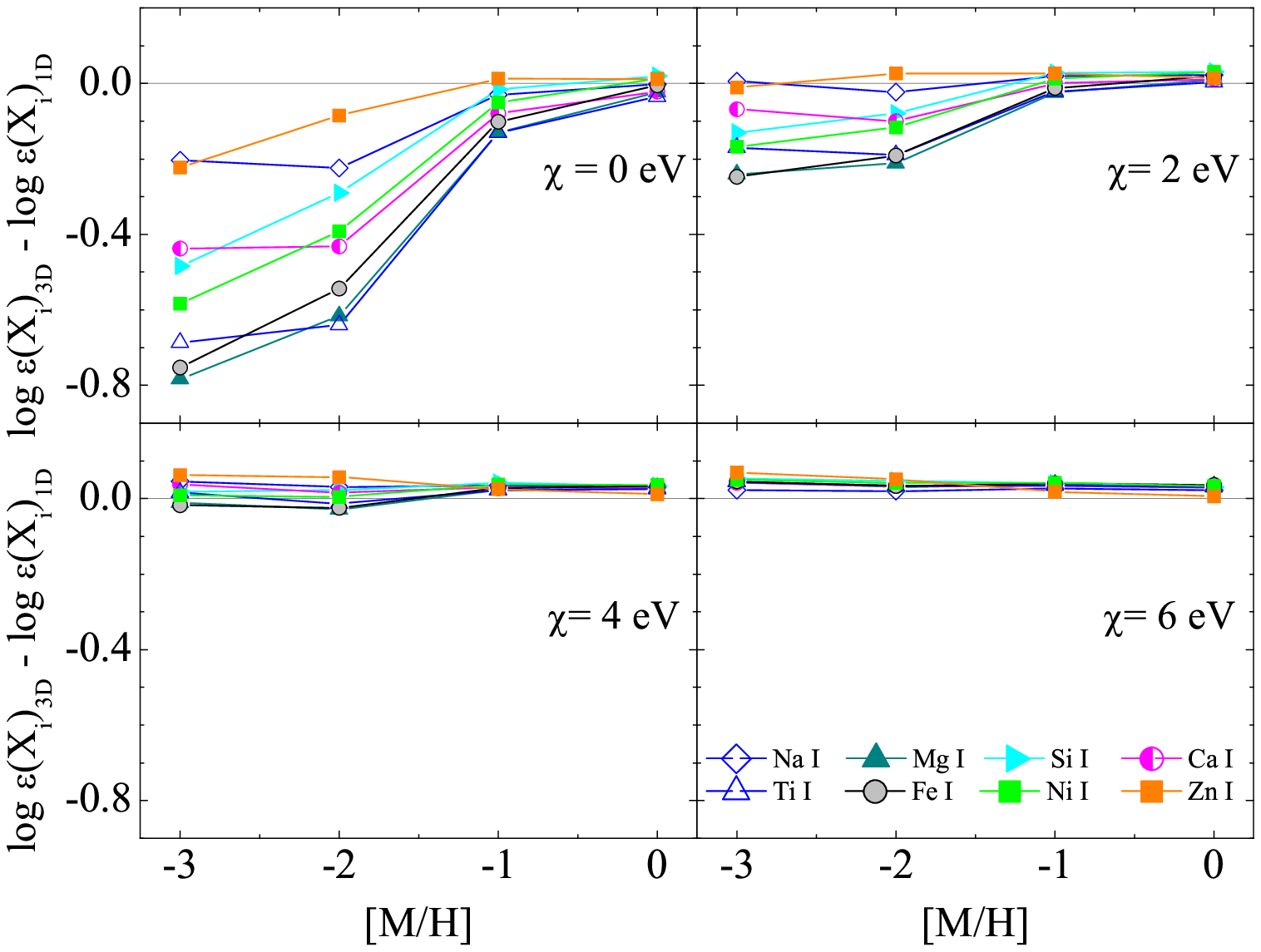}
 \caption{Same as in Fig.2 but at \(\lambda\) = 8500 \AA.}
 \label{Fig3s}
 \end{minipage}
\end{figure}

The smallest 3D--1D abundance corrections in the absolute sense were observed for the highest excitation potentials (Fig. 2 and 3). This is because spectral lines with higher excitation potential form in deeper layers of the stellar atmosphere, where differences between the 3D and 1D temperature stratifications and the amplitude of the horizontal temperature fluctuations are smaller (Fig.~1).

At solar metallicity the slightly hotter mean temperature of the 3D atmosphere (Fig. 1) reduces the line strength of the neutral atoms and horizontal temperature fluctuations increase the line strength [8]. These effects make opposite contributions of approximately equal magnitude to the line strength in the 3D model, thus producing small 3D--1D abundance corrections. However, at low metallicity ([M/H] = --3) the relative importance of horizontal temperature fluctuations is significantly larger which produces stronger lines in 3D and thus results in substantial negative 3D--1D abundance corrections. For neutral atoms, the trends of 3D--1D abundance corrections with metallicity are in general agreement with those obtained by [4], despite slight differences in the atmospheric parameters of the model atmospheres used in both studies.

\subsection{Abundance corrections for ions}

The dependence of 3D--1D abundance corrections on metallicity for singly ionized atoms is shown in Fig. 4 and 5, at \( \lambda \) = 4000 \AA{} and 8500 \AA, respectively. The 3D--1D abundance corrections are significantly lower than in case of neutral atoms and typically do not exceed $\pm$\,0.1\,dex in the metallicity range [M/H] = 0.0 to --3.0 (Fig.~4 and Fig.~5).

\begin{figure}[t]
 \begin{minipage}[t]{0.5\linewidth}
 \centering
 \includegraphics[width=6.9cm]{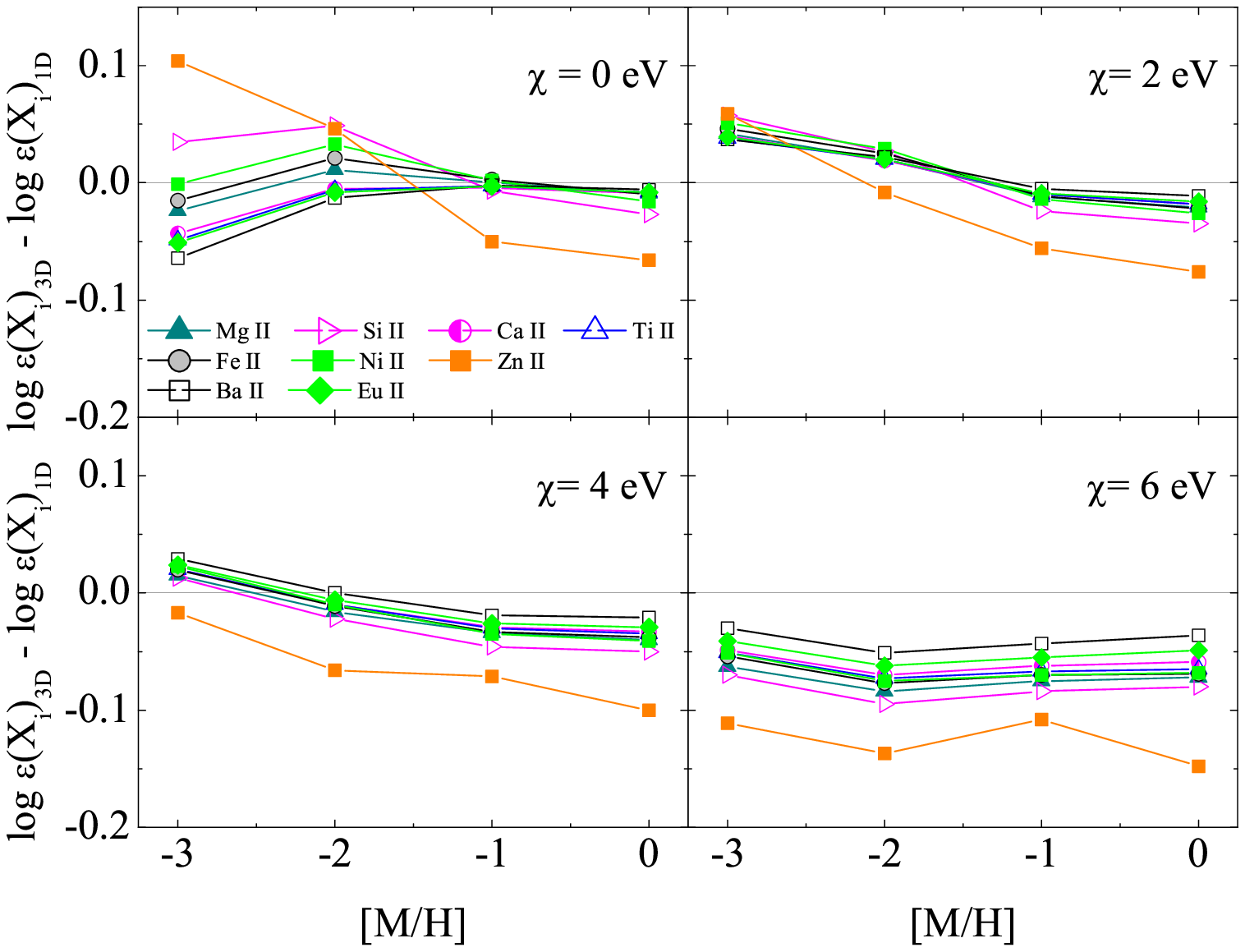}
 \caption{3D--1D abundance corrections for singly ionized atoms plotted as function of metallicity and excitation potential at \(\lambda\) = 4000 \AA.}
 \label{Fig4a}
 \end{minipage}
\hspace{0.0cm}
 \begin{minipage}[t]{0.5\linewidth}
 \centering
 \includegraphics[width=6.9cm]{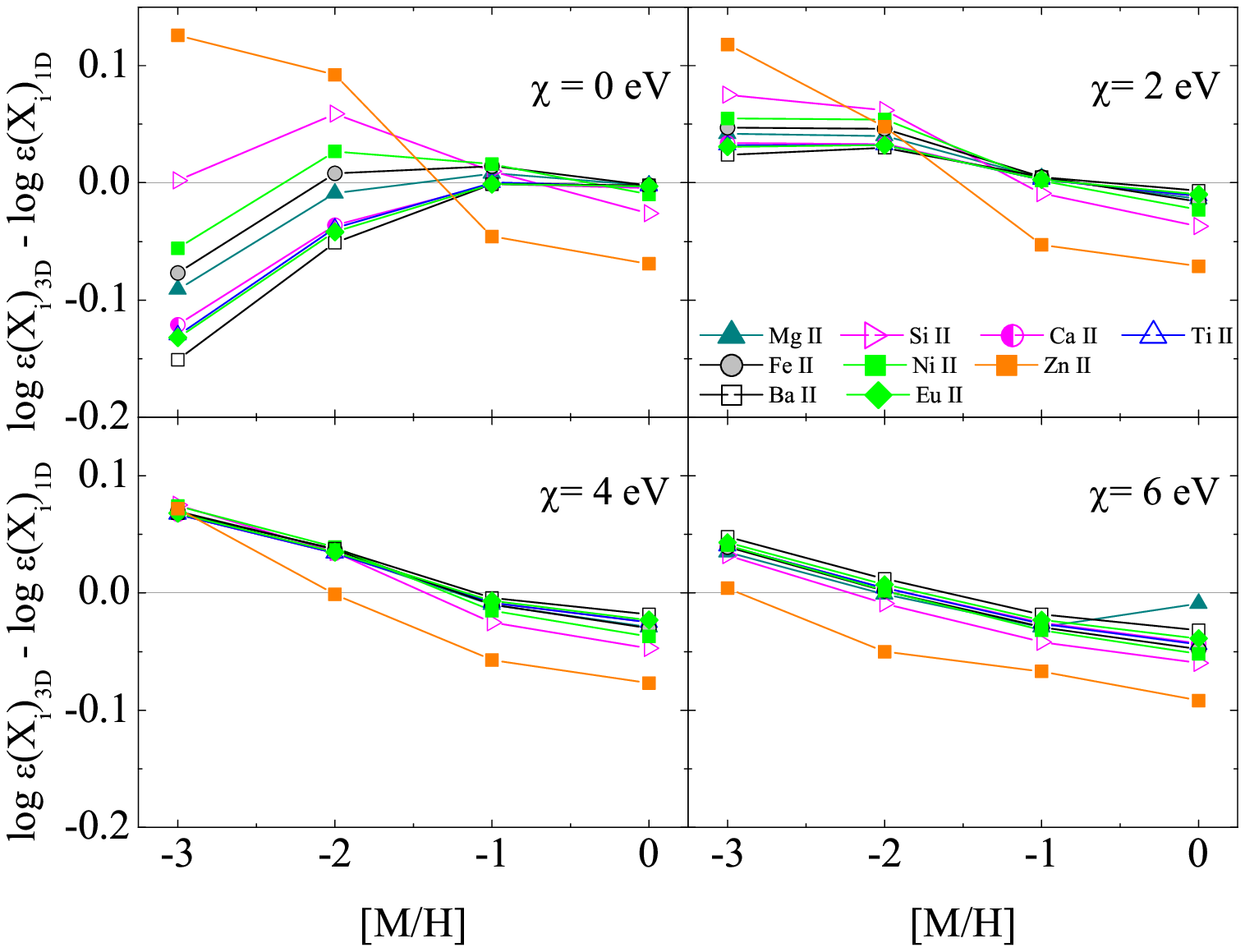}
 \caption{Same as in Fig.4 but at \(\lambda\) = 8500 \AA.}
 \label{Fig5a}
 \end{minipage}
\end{figure}

Temperature inhomogeneities in the deeper atmosphere layers (log \(\tau_\mathrm{Ross}\) \(\approx\) 0) are the main cause of negative 3D--1D abundance corrections for the highest excitation potential lines. On the other hand, low excitation potential lines are mostly influenced by different temperature stratifications predicted by the 3D and 1D models in the outer atmospheric layers, although the differences between the two are small here (Fig. 1).

It is important to note that 3D--1D abundance corrections for ionized atoms in our work become increasingly more negative with increasing excitation potential at fixed metallicity. This is in contrast to the findings obtained by [4].

\section{Conclusions}

We have investigated the influence of convection on the line formation in the photosphere of red giant stars, focusing on the comparison of abundances derived using the 3D hydrodynamical (CO\textsuperscript{5}BOLD) and 1D hydrostatic (LHD) stellar atmosphere models. We have found that the magnitude and sign of the 3D--1D abundance corrections strongly depends on stellar metallicity and atomic parameters of a given spectral line, such as wavelength and excitation potential. At the lowest metallicities ([M/H] < --2.0) the differences between the predictions of the 3D and 1D models may become as large as --0.80 dex for neutral atoms and --0.15 dex for singly ionized atoms. The large difference in abundances predicted by the 3D and 1D models at low metallicities are caused by temporal and horizontal temperature inhomogeneities in the 3D models and different temperature stratification predicted by the 3D and 1D stellar atmosphere models. Impact of the temperature inhomogeneities and lower temperatures in the outer layers of 3D model atmospheres of metal--poor giants in the context of spectral line formation was first noted by [1], later discussed by [4] and confirmed here using a different 3D code. We therefore conclude that if the predictions of currently available 3D stellar atmosphere models are indeed correct these results may signal a warning regarding the usage of 1D stationary stellar atmosphere models in stellar abundance work at low metallicities ([M/H] < --2.0).

The 3D--1D abundance corrections obtained in this work both for neutral and ionized atoms are in general agreement with those derived by [4]. However, the abundance corrections obtained by us are smaller, especially at the lowest metallicity. We argue that significantly larger temperature difference in the outer atmospheric layers between the 3D and 1D model atmospheres used by [4] compared with the models used in this work may be partly responsible for this discrepancy.

In view of these somewhat discrepant findings, we would like to point out that the overall methodology of the two studies is somewhat different since our 3D and 1D models are constructed using the same micro-physics (opacities, equation of state) which is not the case in [4].


\newpage

\begin{thebibliography}{99}
\bibitem{1}
M.~Asplund, \AA. Nordlund, R. Trampedach, R. F. Stein, \emph{3D hydrodynamical model atmospheres of metal-poor stars. Evidence for a low primordial Li abundance}, \emph{A\&A} {\bf 346} L17-L20 (1999).

\bibitem{2}
E.~Caffau, H.-G.~Ludwig, \emph{The forbidden 1082 nm line of sulphur: the photospheric abundance of sulphur in the Sun and 3D effects}, \emph{A\&A}  {\bf 467} L11-L14 (2007).

\bibitem{3}
E.~Caffau, H.-G.~Ludwig, M.~Steffen, B.~Freytag, P.~Bonifacio, \emph{Solar Chemical Abundances Determined with a CO5BOLD 3D Model Atmosphere}, \emph{SoPh} {\bf 268} 255-269 (2011).

\bibitem{4}
R.~Collet, M.~Asplund, R.~Trampedach, \emph{Three-dimensional hydrodynamical simulations of surface convection in red giant stars. Impact on spectral line formation and abundance analysis}, \emph{A\&A} {\bf 469} 687-706 (2007).

\bibitem{5}
B.~Freytag, M.~Steffen, B.~Dorch, \emph{Spots on the surface of Betelgeuse -- Results from new 3D stellar convection models}, \emph{AN} {\bf 323} 213-219 (2002).

\bibitem{6}
B.~Freytag, M.~Steffen, H.-G.~Ludwig, S.~Wedemeyer-B{\"o}hm, W.~Schaffenberger, O.~Steiner, \emph{Simulations of stellar convection with CO5BOLD}, \emph{Journal of Computational Physics: special topical issue on computational plasma physics, ed. Barry Koren} (2011).

\bibitem{7}
H.-G.~Ludwig, E.~Caffau, M.~Steffen, B.~Freytag, P.~Bonifacio, A.~Ku\v{c}inskas, \emph{The CIFIST 3D model atmosphere grid}, \emph{MmSAI} {\bf 80} 711-714 (2009).

\bibitem{8}
M.~Steffen, H.~Holweger, \emph{Line formation in convective stellar atmospheres. I. Granulation corrections for solar photospheric abundances}, \emph{A\&A} {\bf 387} 258-270 (2002).


\end{thebibliography}
\end{document}